\documentclass[12pt]{IOPART}
\usepackage{graphicx}

\def\hc2{$H_{c2}$}

\def\cuscn{$\kappa$-(BEDT-TTF)$_2$Cu(NCS)$_2$}

\def\nh4{$\alpha$-(BEDT-TTF)$_2$NH$_4$Hg(SCN)$_4$}
\newcommand{\gtsim}{\mbox{{\raisebox{-0.4ex}{$\stackrel{>}{{\scriptstyle\sim}}
$}}}}
\newcommand{\ltsim}{\mbox{{\raisebox{-0.4ex}{$\stackrel{<}{{\scriptstyle\sim}}
$}}}}

\begin{document}

\title[The superconducting transition in organic metals]
{A statistical model for the
intrinsically broad superconducting to normal
transition in quasi-two-dimensional crystalline organic metals}
\author{J Singleton$^{a,b,*}$,\footnote[3]{To
whom correspondence should be addressed (j.singleton1@physics.ox.ac.uk)}
N~Harrison$^a$,
CH~Mielke$^a$, JA~Schlueter$^c$ and AM~Kini$^c$}

\address{$^a$~National High Magnetic Field Laboratory, LANL, MS-E536, Los
Alamos, New Mexico 87545, USA}

\address{$^b$~Department of Physics, University of Oxford, The
Clarendon Laboratory, Parks Road, Oxford OX1 3PU, United Kingdom}

\address{$^c$~Materials Science Division,
Argonne National Laboratory, Illinois 60439, USA}

\begin{abstract}
{\sloppy
Although quasi-two-dimensional organic superconductors such as \cuscn
 ~seem to be very clean systems, with apparent
quasiparticle mean-free paths
of several thousand \AA, the superconducting transition
is intrinsically broad (e.g $\sim 1$~K wide for
$T_{\rm c} \approx 10$~K).
We propose that this is due to the extreme anisotropy of these materials, which
greatly exacerbates the statistical effects of
spatial variations in the potential experienced by the quasiparticles.
Using a statistical model, we are able to account for the experimental
observations.
A parameter $\bar{x}$, which characterises
the spatial potential variations, may be derived from Shubnikov-de Haas
oscillation experiments.
Using this value, we are able to predict
a transition width which is in good agreement
with that observed in MHz penetration-depth measurements
on the same sample.}
\end{abstract}

\submitto{\JPCM}

\maketitle
The large number of Shubnikov-de Haas and
de Haas-van Alphen oscillation experiments which have
been carried out on crystalline organic superconductors
demonstrates the high quality of
these materials~\cite{review}; oscillations are
resolved down to $\sim 2$~T~\cite{wosnitza}, and apparent scattering
times extracted by Dingle analysis~\cite{review,shoenberg}
are well in excess of a picosecond, suggesting intralayer mean-free
paths $\gtsim 1000$~\AA~\cite{goddard}. Support for the cleanliness
of the organics also comes from magneto-optical measurements of
cyclotron- and
Fermi-surface-traversal resonances, which again
yield apparent scattering times $\gtsim 1$~ps~\cite{goddard,schrama}.

\begin{figure}[htbp]
\centering
\includegraphics[height=10cm]{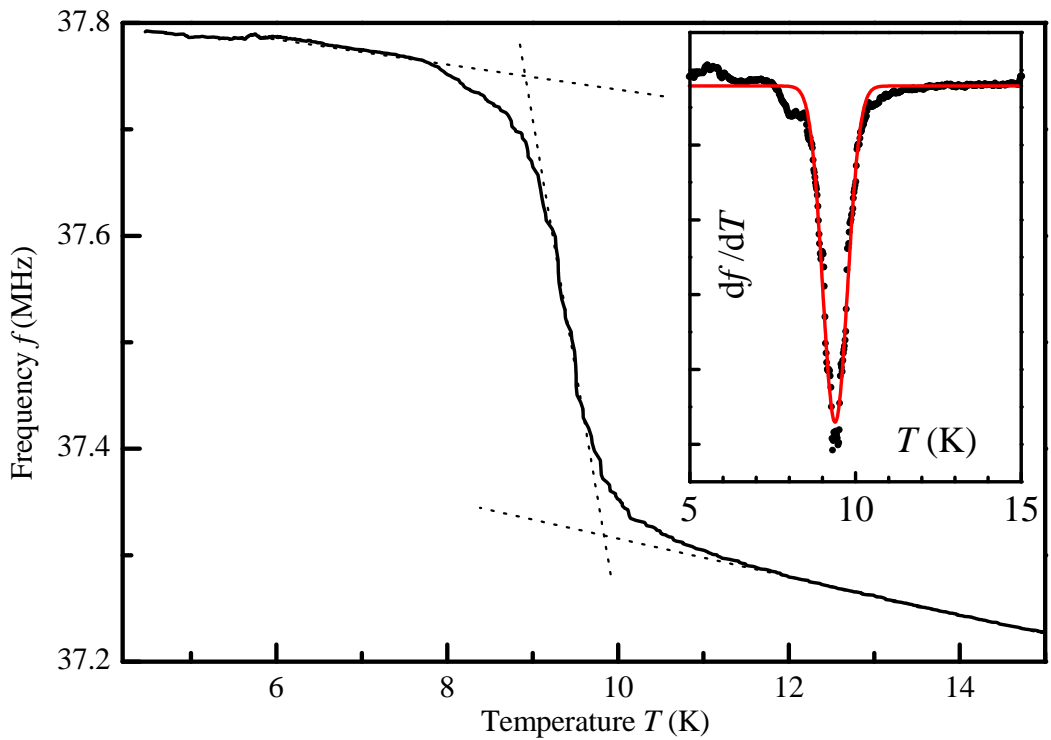}
\caption{MHz penetration data for a single crystal
of \cuscn , shown as resonant frequency $f$ versus
temperature $T$. The superconducting
transition is the steeply sloping region between the more
gentle variations characteristic of superconductivity (low $T$)
and the normal state (high $T$); note
that the complete transition region
occupies a temperature range from
around 8~K to 11~K.
The dotted lines are extrapolations
of the normal-state, transition-region and superconducting-state behaviour.
The intersections of the extrapolations occur at 8.9~T
and 9.8~T, giving $T_{\rm c}^{\rm linear} \approx 9.35$~K (midpoint)
and $\Delta T_{\rm c}^{\rm linear} \approx 0.9$~K.
The inset shows the differential
${\rm d}f/{\rm d}T$ of the data (points) fitted
to a Gaussian (curve) centered on $T_{\rm c}^{\rm Gauss}=9.38$~K,
with a full width of $\Delta T_{\rm c}^{\rm Gauss} \approx 0.7$~K.
}
\label{chuck}
\end{figure}

In spite of this, the superconducting transition
at the critical temperature $T_{\rm c}$
seems to be rather broad, whatever the measurement
method used.
Resistivity measurements are the most prone to
complications~\cite{belin,review2},
especially in an applied magnetic field.
However, even when reliable perturbative techniques such as
thermal conductivity~\cite{belin} or
GHz and MHz penetration-depth experiments~\cite{belin,carrington,mielke1}
are employed in zero field,
the transition has a significant width $\Delta T_{\rm c}$.
Figure~\ref{chuck} is a typical example; a \cuscn ~sample was placed
in a coil forming part of a tank circuit oscillating at
around $38$~MHz (see Ref.~\cite{mielke1}).
The superconducting to normal transition is
observed because of the change from skin-depth
to penetration-depth limited coupling of the sample to the MHz
fields~\cite{carrington,mielke1}, which results in a shift in
resonant frequency $f$ of the tank circuit.
For the purpose of making quantitative
comparisons below, we choose two methods
for defining the temperature $(T)$
width of the transition.
Firstly, by fitting the differential
${\rm d}f/{\rm d}T$ of the data
to a Gaussian centered on $T_{\rm c}^{\rm Gauss}=9.38$~K
(Figure~\ref{chuck}, inset),
a full width of $\Delta T_{\rm c}^{\rm Gauss} \approx 0.7$~K
is obtained.
Alternatively, $\Delta T_{\rm c}$ may be defined using
straight-line extrapolations (see Figure~\ref{chuck}),
to give $\Delta T_{\rm c}^{\rm linear}\approx 0.9$~K.
Note that both methods yield $\Delta T_{\rm c}/T_{\rm c} \sim 0.1$;
it is likely that this significant
intrinsic broadening of the
transition region is responsible for the wide range
of $T_{\rm c}$ values quoted for \cuscn ~\cite{review,review2}.

In this Letter, we show how a superconducting transition can be
broadened due to charged impurities (or vacancies) randomly dispersed
throughout a crystal, even when the potential associated with the
impurity (or vacancies) is of extremely short range. When averaged over
the coherence volume
$\Lambda_{\rm v}=\pi\xi_{0 x}\xi_{0 y}\xi_{0 z}/6$, where the
$\xi$ are Pippard coherence lengths,\footnote{As the following paragraph will
show, we are dealing with a situation in which mean-field theory
breaks down because of spatial variations of the order parameter.
It is therefore inappropriate to use the
Ginzburg-Landau temperature dependent coherence lengths to parameterise
the coherence volume, as these diverge at $T_{\rm c}$.} statistical
variations in the density of impurities (or vacancies) lead to a spatially
varying order parameter $\Delta_0$ and consequently a
Gaussian-broadened transition. The effects of these statistical
variations are shown to become more pronounced when the
dimensionality of the superconductor is reduced, as
is the case in quasi-two-dimensional crystalline organic
superconductors such as
$\kappa$-(BEDT-TTF)$_2$Cu(NCS)$_2$.

Our treatment is closely related to the Ginzburg
criterion~\cite{ginzberg,chaikin},
a quantitative guide to the circumstances under which
mean-field theory can be expected to break down due to fluctuations.\footnote{A
useful introduction to the Ginzburg criterion is given in Section 5.1
of Reference~\cite{chaikin}.}
The Ginzburg criterion has in the past been invoked to explain the
broad superconducting transitions in the
``high $T_{\rm c}$'' cuprates~\cite{cuprates}.
However, in contrast to the situation in the
cuprates, details of the bandstructures of the
organic superconductors are often known to great precision~\cite{review},
whereas the interpretation of heat capacity data (necessary to derive
the Ginzburg criterion) in the latter systems
is still somewhat contentious~\cite{review2}.
We have therefore used an alternative statistical method
to treat the spatial variations of the potential,
based on a good quantitative knowledge of $g(E_{\rm F})$,
the density of quasiparticle states close to the Fermi energy
$E_{\rm F}$.

The introduction of impurities in an ideal metal leads to
potential variations (of typical length scale $R$)
and a finite scattering rate $\tau^{-1}$
for {\it normal} quasiparticles~\cite{review,alloy}.
By contrast, the spatial extent $\xi_0$ of the {\it superconducting} wavefunction
often greatly
exceeds $R$, so that theoretical studies
fail to yield a direct correlation between $\tau^{-1}$
and the order parameter $\Delta_0$~\cite{parks}.\footnote{Except in
the case of magnetic impurities~\cite{parks}.}
Nevertheless, nonmagnetic impurities can have an effect because
the superconducting state is sensitive to changes
in $g(E_{\rm F})$.
As an example of this, let us consider the weak-coupling BCS formula
\begin{equation}\label{BCS}
\Delta_0\approx\hbar\omega_0\exp\bigg[-\frac{1}{g(E_{\rm F})V}\bigg],
\end{equation}
where $V$ is an interaction strength~\cite{parks}.
If a small (local) fraction $x$ of the host atoms or molecules are
replaced by impurities, then
the variation of $g(E_{\rm F},x)$ can be written~\cite{alloy}
\begin{equation}\label{density}
    g(E_{\rm F},x)=g(E_{\rm F})+g^\prime(E_{\rm F})x.
\end{equation}
If we assume that each impurity
introduces one extra charge ($-e$),
the derivative $g^{\prime}(E_{\rm F})=({\rm d} g(E_{\rm F},x)/{\rm d} x)$
can be obtained from bandstructure calculations
in the limit $x \rightarrow 0$.

The number of sites available for impurity
substitution
within $\Lambda_{\rm v}$ is $n=\Lambda_{\rm v}/u_{\rm v}$,
where $u_{\rm v}$ is the formula-unit volume.
However, the {\it local} number of impurities within such a volume,
$m=xn$, will be subject to
statistical variations via the Binomial distribution (BD)~\cite{alloy}
\begin{equation}\label{binomial}
    p(m,n)=\frac{\bar{x}[1-\bar{x}]^{n-m}n!}{m![n-m]!},
\end{equation}
where $\bar{x}$ is the mean of $x$.
Below we consider relatively large values of $n$ ($43 \ltsim n \ltsim 10^{10}$);
the skewness factor $\eta=1/\sqrt{6\bar{x}[1-\bar{x}]n}$
vanishes for large $n$,
and the BD is well
approximated by the normal distribution~\cite{alloy}.
Hence, the mean value of $m$ becomes $n\bar{x}$ while its standard
deviation is $\sigma(m)=\sqrt{n\bar{x}[1-\bar{x}]}$. In
a ``clean'' metal
$\bar{x}\ll 1$, leading
to a standard deviation in $x$
of
\begin{equation}
\sigma(x)\approx\sqrt{\bar{x}/n}.
\label{tedium}
\end{equation}
Using the fact that the standard
deviation of a function of $\bar{x}$ is equal to the derivative of
that function multiplied by the standard deviation of $\bar{x}$,
we insert Equations~\ref{density} and \ref{tedium}
into Equation~\ref{BCS}, yielding
the standard deviation $\sigma(\Delta_0)$ of $\Delta_0$,
\begin{equation}\label{result}
    \frac{\sigma(\Delta_0)}{\Delta_0}\approx
    \frac{g^\prime(E_{\rm F})}{g(E_{\rm F})^2V}\sqrt{\frac{\bar{x}u_{\rm
    v}}{\Lambda_{\rm v}}}.
\end{equation}
Applying the relation $2\Delta=\alpha k_{\rm B}T_{\rm c}$,
where $\alpha \approx 3.52$ in the weak-coupling BCS limit~\cite{parks},
the broadening of the superconducting transition is
\begin{equation}\label{broadening}
    \frac{\Delta T_{\rm c}}{T_{\rm c}}\equiv
    2\frac{\sigma (T_{\rm c})}{T_{\rm c}}\approx
    2\frac{\sigma(\Delta_0)}{\Delta_0}.
\end{equation}
We now demonstrate the sensitivity of $\Delta T_{\rm c}$ to
the dimensionality of the superconductor.

{\it Conventional three-dimensional (3D) superconductors}
have coherence lengths $\sim 10^2-10^4$~\AA
~and $u_{\rm v}\sim$~20~\AA$^3$~\cite{ashcroft},
yielding $10^4 \ltsim n \ltsim 10^{10}$.
In reasonably pure metals, we expect
$10^{-4} \ltsim \bar{x} \ltsim 10^{-2}$, while $g(E_{\rm F})V \sim 0.3$;
finally we use the fact that
$g^\prime(E_{\rm F})/g(E_{\rm F})=\frac{1}{3}$ for a free-electron
model~\cite{ashcroft}.
These figures yield
transition widths $10^{-3} \ltsim \Delta T_{\rm c}/T_{\rm c} \ltsim 10^{-7}$,
in reasonable agreement with observations~\cite{parks}.
These sharp transitions
are a consequence of the large size of the
superconducting wavefunction, allowing inhomogeneities
to be averaged out.

In {\it quasi-two-dimensional (Q2D) organic superconductors},
the intralayer coherence lengths are $\sim 10^2$~\AA.
However, the {\it inter-layer} coherence length $\xi_{0z}$
can be much less than the layer spacing $a$~\cite{review2};
in $\kappa$-(BEDT-TTF)$_2$Cu(NCS)$_4$, $\xi_{0 z}$ can be
estimated to be $\sim 0.3$~\AA,\footnote{The anisotropy
is $\sim 100-350$~\cite{carrington,mansky,pratt}, and the
in-plane coherence lengths are estimated to be
$\approx 76$~\AA~\cite{review2,ishiguro}.}
whereas $a\approx 16$~\AA~\cite{ishiguro}.
The consequence of this extreme
anisotropy is that $\Lambda_{\rm v}$ is replaced by
a ``coherence area''
$\Lambda_{\rm a}=\pi\xi_{0 x}\xi_{0 y}/4$, because
superconducting wavefunctions do not extend out of the
layers. In enumerating $n$, the unit cell must also
be represented by an area $u_{\rm a} (\approx 106$~\AA$^2$ in
$\kappa$-(BEDT-TTF)$_2$Cu(NCS)$_4$~\cite{ishiguro}). Assuming
BCS values for the intralayer coherence lengths
($\xi_{0 x} \approx\xi_{0 y} \approx 76$~\AA)~\cite{ishiguro},
we obtain $n \approx 43$, much less than typical values
of $n$ in 3D systems.

In the current context,
\cuscn ~has the considerable advantage that its
intralayer bandstructure
may be represented to good accuracy
by the {\it effective dimer model}~\cite{goddard,caulfield,schmalian}
\begin{equation}
E({\bf k})=
\pm 2\cos(\frac{k_{\bf b}b}{2})
\sqrt{t_{{\bf c}1}^2+t_{{\bf c}2}^2+2t_{{\bf c}1}t_{{\bf c}2}\cos(k_{\bf
c}c)}
+2t_{\bf b}\cos(k_{\bf b}b).
\label{phwoar}
\end{equation}
Here $k_{\bf b}$ and $k_{\bf c}$ are the intralayer components of {\bf k}
and $t_{\bf b}$, $t_{{\bf c}1}$
and $t_{{\bf c}2}$ are interdimer transfer
integrals~\cite{caulfield,schmalian};
the $+$ and $-$ signs result in the quasi-one-dimensional (Q1D) sheets and Q2D pocket
of the Fermi surface respectively~\cite{review}.
Moreover, the Fermi-surface warping in the interlayer direction is rather small~\cite{goddard},
so that it may be ignored for the current purposes.
Accurate de Haas-van Alphen and
magnetic breakdown data constrain the parameters in Equation~\ref{phwoar}
rather tightly, leading to values $t_{\bf b}=15.6$~meV, $t_{{\bf c}1}=24.2$~meV and
$t_{{\bf c}2}=20.3$~meV~\cite{goddard}.

\begin{figure}[htbp]
\centering
\includegraphics[height=10cm]{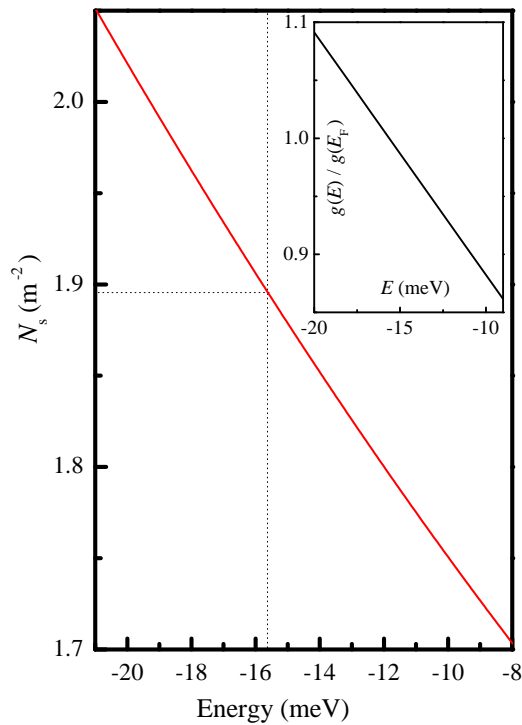}
\caption{$N_{\rm s}$, the areal quasiparticle density per layer,
as a function of quasiparticle energy $E$ for \cuscn ; the curve has been
derived using the parameters listed in the text and Equation~\ref{phwoar}.
The position of the Fermi energy and the value of $N_{\rm s}$ corresponding to the
well-known magnetic breakdown $\beta$ frequency~\cite{review} are indicated by dotted lines.
The inset shows the variation of the quasiparticle density of
states $g(E)$, normalised to its value at $E=E_{\rm F}$,
as a function of quasiparticle energy.
}
\label{fig2}
\end{figure}

Figure~\ref{fig2} shows $N_{\rm s}$, the areal quasiparticle density per layer,
as a function of quasiparticle energy $E$ for \cuscn ; the curve has been
derived using the parameters listed in the previous paragraph and Equation~\ref{phwoar}.
The position of the Fermi energy and the value of $N_{\rm s}$ corresponding to the
well-known magnetic breakdown $\beta$ frequency~\cite{review},
which encompasses the entire Fermi surface, are indicated by dotted lines.
Whereas for many purposes~\cite{goddard} the bands of \cuscn are rather parabolic
close to $E_{\rm F}$, some curvature of $N_{\rm s}$ versus $E$ is visible in Figure~\ref{fig2},
resulting in a $g(E_{\rm F})$ which varies with energy (Figure~\ref{fig2}, inset).
Using the results shown in Figure~\ref{fig2},
we obtain $g^\prime(E_{\rm F})/g(E_{\rm F})\approx 1.45$.

Whilst it is unlikely that \cuscn ~is a weak-coupling BCS
superconductor~\cite{review,carrington,schmalian,elsinger,french},
Equation~\ref{BCS} is known to describe the {\it functional
dependence} of $T_{\rm c}$ on $g(E_{\rm F})$ very well,
as shown in the pressure-dependent experiments of Caulfield {\it et al.}~\cite{caulfield}.
Using $g(E_{\rm F})V \approx 0.3$~\cite{caulfield}
and $10^{-4} \ltsim \bar{x} \ltsim 10^{-2}$ we obtain
$0.015 \ltsim \Delta T_{\rm c}/T_{\rm c}\ltsim 0.15$,
encompassing all known experimental data.

In order to find a suitable value of $\bar{x}$ for the
sample used in Figure~\ref{chuck}, we turn to Reference~\cite{alloy},
which shows that spatial variations in the potential
experienced by the quasiparticles leads to broadening
of Landau levels and hence damping of Shubnikov-de Haas
and de Haas-van Alphen oscillations. This damping
is parameterised by an effective Dingle temperature
\begin{equation}\label{dingle}
    T_{\rm D}=\frac{\bar{x}[1-\bar{x}]F^\prime(\bar{x})^2a}
    {\pi k_{\rm B}m^\ast}\sqrt{\frac{\hbar e^3}{2F}}.
\end{equation}
In the current context,
$F$ is the quantum oscillation frequency of the
$\beta$ breakdown orbit and $F^\prime = {\rm d}F/{\rm d}x$;
it is simple to show that $F^\prime\equiv F\approx$~3920~T~\cite{review}.
The sample used in Figure~\ref{chuck} had
$T_{\rm D \alpha} \approx 0.42$~K~\cite{goddard}
for the $\alpha$-orbit Shubnikov-de Haas oscillations,
which translates into a Dingle temperature of $T_{\rm D} \approx 0.76$~K
for the $\beta$
magnetic breakdown orbit~\cite{review}.
Substituting this value of $T_{\rm D}$ into Equation~\ref{dingle}
yields $\bar{x} \approx 0.0017$,
which can be used in Equation~\ref{broadening} to give
$\Delta T_{\rm c} \approx 0.6$~K.
This value is in good agreement with
$\Delta T_{\rm c}^{\rm Gauss} \approx 0.7$~K extracted from the MHz
experiments (see Figure~\ref{chuck}, inset) and close to the
value $\Delta T_{\rm c}^{\rm linear} \approx 0.9$~K found using the
alternative linear extrapolation method.

One possible interpretation of the value of $\bar{x} \approx 0.0017$
is that
$\sim 0.1-0.2$\% of the molecular sites in our crystals of
\cuscn ~are in some way defective, functioning as
``impurities'' or ``vacancies''; possible mechanisms
might include neutral BEDT-TTF molecules, anions
which are missing or in which the Cu ion possesses the
wrong charge, or the incorporation of other molecular
species from the growth process. The detection of such
defects by other means is very difficult; all that
can be noted at present is that such concentrations
of defects are thought to be not unlikely~\cite{day}.

In summary, using a statistical model, we are able to account for the
broadened superconducting to normal transitions observed in
organic superconductors.
Our model consistently explains the
superconducting transition width and the Landau-level broadening
in $\kappa$-(BEDT-TTF)$_2$Cu(NCS)$_4$ using one
parameter, $\bar{x}$.
Given the precise knowledge of the Fermi-surface topologies
of many organic superconductors, and the availability of
many samples of differing quality and effective dimensionality~\cite{mielke1},
our model may be useful in obtaining a more detailed understanding of
the factors which influence the formation of the superconducting state.

The work is supported by the Department of Energy, the National
Science Foundation (NSF) and the State of Florida (USA), and
by EPSRC (UK). We should like to thank Ross McKenzie, Steve Blundell,
Albert Migliori, Bill Hayes and Jochen Wosnitza for useful comments.
Dieter Schweitzer, Peter Day, Urs Geiser and Mike Montgomery
are thanked for discussions about the likely number of defects
present in charge-transfer salts.


\vspace{2cm}

\begin{thebibliography}{99}
\bibitem{review}
J. Singleton, {\it Rep. Progr. Phys.} {\bf 63}, 1111 (2000).
\bibitem{wosnitza}
See e.g. J. Wosnitza, S. Wanka,
R. Haussler, H. v. Lohneysen, J.A. Schlueter,
U. Geiser, P.G. Nixon, R.W. Winter and G.L. Gard, Phys. Rev. B
{\bf 62}, 11973 (2000).
\bibitem{shoenberg}
David Shoenberg, {\it Magnetic oscillations in metals}
(Cambridge University Press, 1982).
\bibitem{goddard}
J.~Singleton, P.A. Goddard, A. Ardavan,
N.~Harrison, S.J. Blundell, J.A.~Schlueter and A.M.~Kini,
preprint cond-mat 0104570,
submitted to Phys. Rev. Lett.
\bibitem{schrama}
J.M.~Schrama, J. Singleton, R.S. Edwards,
A. Ardavan, E. Rzepniewski, R. Harris, P. Goy,
M. Gross, J. Schlueter, M. Kurmoo and P. Day,
J. Phys.: Condens. Matter {\bf 13} 2235 (2001).
\bibitem{belin}
S.~Belin, T.~Shibauchi, K.~Behnia and T.~Tamegai,
J. of Superconductivity, {\bf 12}, 497 (1999).
\bibitem{review2}
J. Singleton and C.H. Mielke, {\it Contemp. Phys.},
in press.
\bibitem{carrington}
A. Carrington,
I.J. Bonalde, R. Prozorov, R.W. Gianetta, A.M. Kini,
J. Schlueter, H.H. Wang, U. Geiser and J.M. Williams,
Phys. Rev. Lett., {\bf 83},
4172 (1999).
\bibitem{mielke1}
Charles~Mielke, John Singleton, Moon-Sun Nam,
Neil Harrison, C.C. Agosta, B. Fravel and L. K. Montgomery,
J. Phys.: Condens. Matt. {\bf 13}, 8325 (2001).
\bibitem{ginzberg}
V.L. Ginzburg, Fiz. Tverd. Tela. {\bf 2}, 2031 (1960);
Sov. Phys. Solid State {\bf 2}, 1824 (1961).
\bibitem{chaikin}
P.M. Chaikin and T.C. Lubensky,
{\it Principles of condensed matter physics}
(Cambridge University Press, 1995).
\bibitem{cuprates}
S.E. Inderhees, M.B. Salamon, N. Goldenfeld, J.P. Rice, B.G. Pazol,
D,M. Ginsberg, J.Z. Liu and G.W. Crabtree,
Phys. Rev. Lett. {\bf 60}, 1178 (1988).
\bibitem{alloy}
N. Harrison and J. Singleton, J. Phys.: Condens. Matt.
{\bf 13}, L463 (2001).
\bibitem{ashcroft}
N.W.~Ashcroft and N.D.~Mermin, {\it Solid State Physics}, Saunders
(1976).
\bibitem{parks}
{\it Superconductivity}, edited by R.D.~Parks
(Marcel Dekker, New York, 1969).
\bibitem{mansky}
P.~A.~Mansky, P.~M.~Chaikin, and R.~C.~Haddon, Phys. Rev. B
{\bf 50}, 15929 (1994).
\bibitem{pratt}
F.L. Pratt, S.L. Lee, C.M. Aergerter, C. Ager, S.H. Lloyd,
S.J. Blundell, F.Y. Ogrin, E.M. Forgan, H. Keller, W. Hayes,
T. Sasaki, N. Toyota and S. Endo, Synth. Met. {\bf 120},
1015 (2001); F.L. Pratt {\it et al.}, preprint.
\bibitem{ishiguro}
T. Ishiguo, K. Yamaji and G. Saito. {\it Organic Superconductors}
(Springer-Verlag, Berlin 1998).
\bibitem{caulfield}
J.M.~Caulfield, W.~Lubczynski, F.L.~Pratt,
J.~Singleton, D.Y.K.~Ko,
W.~Hayes, M.~Kurmoo and P.~Day,
J. Phys.: Condens. Matter {\bf 6}, 2911 (1994).
\bibitem{schmalian}
J.~Schmalian, {\it Phys. Rev. Lett.} {\bf 81}, 4232 (1998).
\bibitem{elsinger}
H. Elsinger, J. Wosnitza, S. Wanka, J. Hagel,
D. Schweitzer and W. Strunz, Phys. Rev. Lett. {\bf 84}, 6098 (2000).
\bibitem{french}
S. Lefebvre, P. Wzietek, S. Brown, C. Bourbonnais, D. Jerome, C. Meziere,
M. Fourmigue and P. Batail, Phys. Rev. Lett. {\bf 85}, 5420 (2000).
\bibitem{day}
P. Day, D. Schweitzer, L.K. Montgomery and U. Geiser, private discussions
(2000-2001).

\end{thebibliography}
\end{document}